\documentstyle[11pt,dunk2001_asp,twoside,epsf,psfig]{article}
\def\sech{{\rm \;sech}}

\def\O{\over}

\def\subr #1{_{{\rm #1}}}

\def\edcomment#1{\iffalse\marginpar{\raggedright\sl#1\/}\else\relax\fi}
\reversemarginpar

\begin{document}
\title{Structure and dynamics of disks in galaxies}
 \author{P.C. van der Kruit}
\affil{Kapteyn Astronomical Institute, University of Groningen, P.O.
Box 800, 9700 AV Groningen, the Netherlands, vdkruit@astro.rug.nl}

\begin{abstract}
One the most cited papers in astronomy is Ken Freeman's 1970 paper on
exponential disks in galaxies. In this contribution I review what has
been done in this area since then and what we can infer about 
systematic properties of disks in galaxies from surface photometry, HI
synthesis observations and measurements of stellar kinematics. Most disks
have radial truncations at $3.6 \pm 0.6$ radial scalelengths $h\subr{R}$. 
Galaxy disk thicknesses $h\subr{R}/h\subr{z}= 7.3 \pm 2.2$ imply that these 
disk cannot be ``maximum disks''. I briefly 
discuss a recent study of the ``superthin'' edge-on galaxy IC 5249.
\end{abstract}

\section{Introduction}

This meeting is a tribute to Ken Freeman, his research and that with his
students. Ken's most cited paper is undoubtedly {\it ``On the disks of
spiral and S0 galaxies''} (Freeman, 1970).
Although it was de Vaucouleurs (e.g. 1959), who first showed
observationally that the surface brightness distributions can be
described by an exponential law, it was Ken in his seminal paper who
collected the material available and studied the distribution of
parameters and the dynamical relations. He derived the rotation
curve and angular momentum distribution for an exponential disk in
centrifugal equilibrium, found that 28 of the 36 galaxies have approximately 
the same face-on central
surface brightness (known as ``Freeman's law'') and on the basis of the
rotation curves of NGC300 and M33 concluded that there must be
undetected matter of at least the mass of the detected galaxy.

{\small
\begin{table}
\caption{ADS Citations up to end 2000 of important papers}
\begin{tabular}{||l|l l|l|r|r|r||}
\hline
Author(s) & Year & Reference & Subject & \multicolumn{3}{c||}{ADS citations} \\
 & & & &  Tot. &  9/0 & yr$^{-1}$ \\
\hline
Sch\"onberg \& & 1942 & ApJ {\bf 96}, 161 & H-burning and & 57 & 
5 & 1.0 \\
\ \ Chandrasekhar & & & \ \ S-C core & & & \\
Baade & 1944 & ApJ {\bf 100}, 137 & Stellar populations & 129 & 5 & 2.3 \\
de Vaucouleurs & 1948 & An'Ap {\bf 11}, 247 & $R^{1/4}$-law & 388 & 42 & 
7.5 \\
Spitzer \& & 1951 & ApJ {\bf 114}, 385 & Secular evolution of & 111 & 5 & 
2.3 \\
\ \ Schwarzschild &  & & \ \ stellar motions & & & \\
Sandage \& & 1952 & ApJ {\bf 116}, 463 & Giant-branch evolu- & 55 & 6 & 
1.1 \\
\ \ Schwarzschild & & & \ \ tion; MS turn-off & & & \\
Hoyle \& & 1955 & ApJSuppl {\bf 2}, 1 & Giant-branch evolu- & 87 & 5 & 
1.9 \\
\ \ Schwarzschild & & & \ \ tion;  HR-diagrams  & & & \\
Salpeter & 1955 & ApJ {\bf 121}, 161 & Salpeter-function & 922 & 154 & 20.5 \\
Burbidges, & 1957 & RMP {\bf 29}, 547 & Stellar nucleosyn-  & 
343 & 15 & 8.0 \\
\ \ Fowler \&\ Hoyle & & & \ \ thesis & & & \\
Oort, Kerr \& & 1958 & MN {\bf 118}, 379 & HI structure of the & 40 & 2 & 2.0 \\
\ \ Westerhout & & & \ \ Galaxy & & & \\
Schmidt & 1959 & ApJ {\bf 129}, 243 & SF; ``Schmidt-law'' & 376 & 39 & 
9.2 \\
King & 1962 & AJ {\bf 67}, 471 & King law & 440 & 42 & 11.6 \\
Sandage & 1962 & ApJ {\bf 135}, 333 & NGC188 and chem- & 98 & -- & 2.6 \\
 & & & \ \ ical evolution  & & & \\
Eggen, Lynden- & 1962 & ApJ {\bf 136}, 748 & Collapse of the  & 670 & 
61 & 17.6 \\
\ \ Bell \& Sandage & & & \ \ Galaxy & & & \\
Schmidt & 1963 & ApJ {\bf 137}, 758 & G-dwarf problem & 227 & 10 & 6.1 \\
Toomre & 1964 & ApJ {\bf 139}, 1217 & Local disk stability & 607 & 83 & 16.9 \\
Lin \&\ Shu & 1964 & ApJ {\bf 140}, 646 & Density wave theory & 232 & 18 & 
6.4 \\
Goldreich \& & 1965 & MN {\bf 130}, 97 & Disk instability and & 128 & 
14 & 3.7 \\
\ \ Lynden-Bell & 1965 & MN {\bf 130}, 125 & \ \ spiral structure 
& 195 & 27 & 5.6 \\
King & 1966 & AJ {\bf 71}, 64 & King models & 803 & 46 & 23.6 \\
Lynden-Bell & 1967 & MN {\bf 136}, 101 & Violent relaxation & 347 & 25 & 10.5 \\
Schmidt & 1968 & ApJ {\bf 151}, 393 & Quasars; $V/V_{\bf m}$-test & 504
& 34 & 15.8 \\
Tinsley & 1968 &  ApJ {\bf 151}, 547 & Photometric evolu- & 66 & 5 & 2.1 \\
 & & & \ \ tion & & & \\
Freeman & 1970 & ApJ {\bf 160}, 811 & Exponential disks & 820 & 84 & 27.3 \\
Freeman, San- & 1970 & ApJ {\bf 160}, 831 & Origin of Hubble & 232 & 
11 & 7.7 \\
\ \ dage \&\ Stokes & & & \ \ types & & & \\
Searle & 1971 & ApJ {\bf 168}, 327 & Disk abundance & 349 & 10 & 
12.0 \\
 & & & \ \ gradients & & & \\
Peebles & 1971 & A\&A {\bf 11}, 377 & Origin of angular & 54 & 5 & 1.9 \\
 & & & \ \  momentum  & & & \\
Toomre \& Toomre & 1972 & ApJ {\bf 178}, 623 & Interacting galaxies & 785 & 
90 & 28.0 \\
Searle, Sargent \&\ & 1973 & ApJ {\bf 179}, 427 & Photometric evol- & 
337 & 20 & 12.5 \\
\ \ Bagnuolo & & & \ \ ution of galaxies & & & \\
Ostriker \&  & 1973 & ApJ {\bf 186}, 467 & Disk stability and & 439 & 
29 & 16.3 \\
\ \ Peebles & & & \ \ dark halos & & & \\
Tully \&\ Fisher & 1977 & A\&A {\bf 54}, 66 & TF-relation & 547 & 63 & 23.8 \\
Wielen & 1977 & A\&A {\bf 60}, 263 & Secular evolution of & 220 & 29 & 9.6 \\
 & & & \ \ stellar motions & & & \\
Larson \&\ Tinsley & 1978 & ApJ {\bf 219}, 46 & Photometric evolu- & 554
& 29 & 25.2 \\
 & & & \ \ tion & & & \\
Searle \&\ Zinn & 1978 & ApJ {\bf 225}, 357 & Globular clusters & 
539 & 58 & 24.5 \\
 & & & \ \ and halo formation & & & \\
Tinsley & 1980 & FCP {\bf 5}, 287 & Photometric and & 449 & 57 & 15.0 \\
 & & & \ \ chemical evolution & & & \\
van der Kruit \& & 1981 & A\&A {\bf 95}, 105 & 3-D galaxy disk & 280 
& 20 & 14.8 \\
\ \ Searle & & & \ \ model & & & \\
Gilmore \&\ Reid & 1983 & MN {\bf 202}, 1025 & Galactic thick disk & 291 & 
16 & 17.1 \\
\hline
\end{tabular}
\end{table}
}

\begin{figure}
\centerline{\vbox{
\psfig{figure=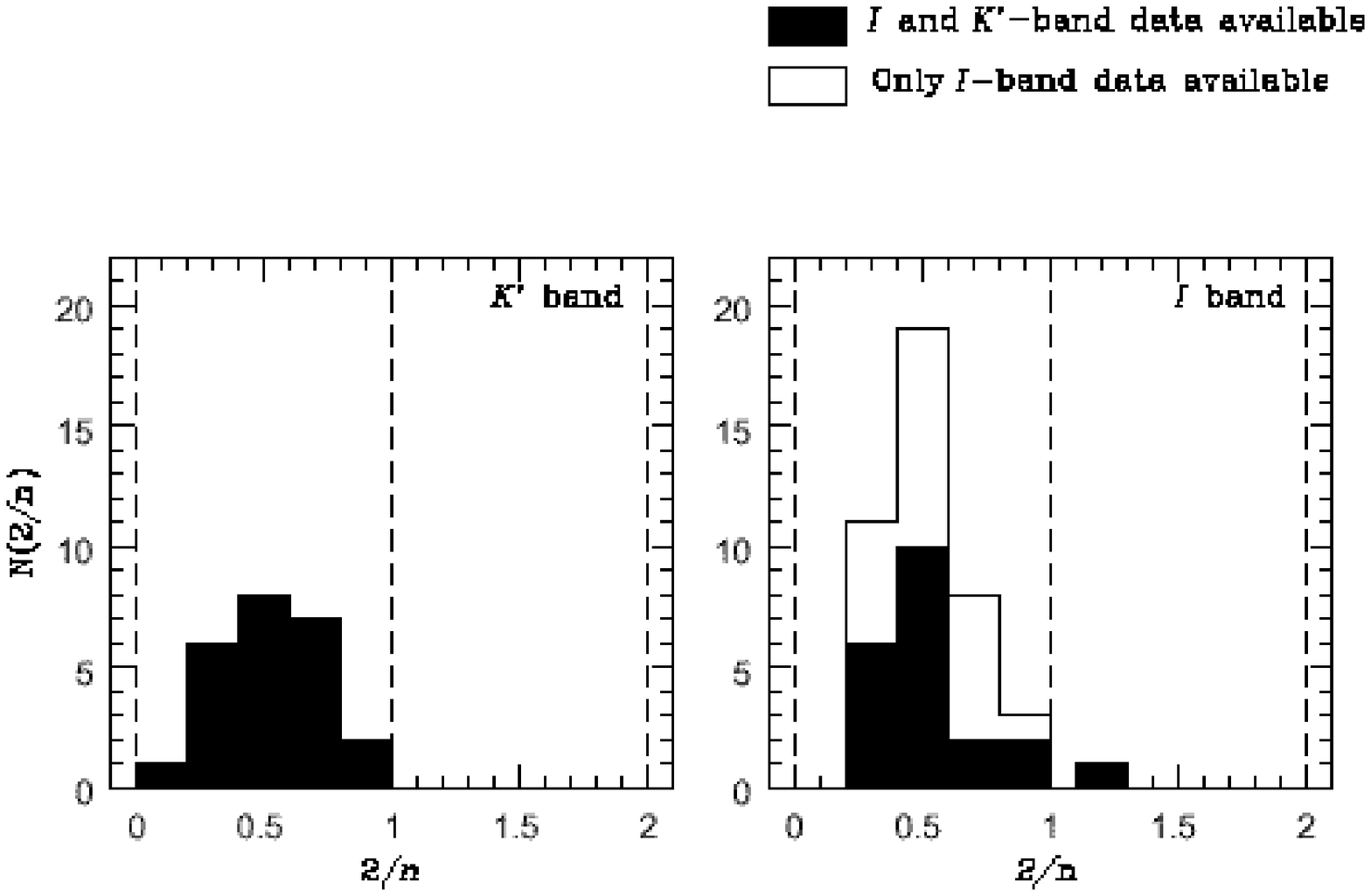,width=8cm}
}}
\caption{The distribution in $K'$ and $I$ of the exponent $2/n$ 
in eq. (2) for a sample of edge-on galaxies (from de Grijs 
{\it et al.}, 1997).}
\end{figure}

It is of some interest to see how Ken's 1970-paper compares in citation rate
to others. For this purpose I collected such information using 
the NASA Astrophysics Data System (ADS). 
The citation scores in ADS are not complete, but certainly
indicative and internally consistent.  The first exercise was to draw 
up a list of what I feel are the most important papers related to
studies of structure of galaxies. In Table 1 I give the total citation
score up to the end of 2000, that for 1999+2000 (``9/0'') and 
the average number of citations per year. In this listing I
find only a few papers that are comparable or exceed Ken's 1970-paper in
citations. These are Ed Salpeter's 1955 paper on star formation in the
Galactic disk in which he defines the ``Salpeter function'', Ivan
King's 1966 paper on the dynamics of globular clusters and the ``King
models'' and the study by the Toomre \&\ Toomre in 1973 on models for
interacting galaxies. 

To complete the search I also checked in ADS which papers 
between 1945 and 1975 were annually the most highly cited papers in Ap.J.
(main journal), A.J., M.N.R.A.S. and A\&A. One that comes close is 
the galaxy redshift survey of Humason {\it et al.} (1956) 
(754 citations, 16 in 1999+2000). Outside the field of galaxies
there are two papers that clearly exceed it, namely Shakura \&\ 
Sunyaev (1973; A\&A {\bf 24}, 337; 1823 citations and 313 in 1999+2000)
on the appearance of black holes in binary systems and Nino Panagia
(1973, AJ {\bf 78}, 929; 1026 and 46) on parameters of early type stars, 
while Brocklehurst (1971, MN {\bf 153}, 471; 787 and 28) on hydrogen
population levels in gaseous nebulae comes close.

In any case, Ken's 1970 paper is well within the absolute top ten of citation
scores of papers in astronomy and deservedly so.

\section{Properties of disks}

The exponential nature of the light distribution in galactic disks was
extended to a three-dimensional model using the self-gravitating 
isothermal sheet description for the vertical distribution 
(van der Kruit \&\ Searle, 1981)\footnote{As explained in that paper, the idea
to use the isothermal sheet was inspired by a remark made by Ken Freeman during
IAU Symposium {\bf 77} in 1978.}:
\begin{equation}
L(R,z) = L(0,0) {\rm \;e}^{-R/h\subr{R}} \sech ^{2} \left( {z \O z_{\circ}} 
\right).
\end{equation}
The isothermal assumption was later dropped and replaced by the family of models
(van der Kruit, 1988) 
\begin{equation}
L(R,z) = L(0,0) {\rm \;e}^{-R/h\subr{R}} \sech ^{2/n} \left( {{n z} \O
{2 h\subr{z}}} \right).
\end{equation}
This ranges from the isothermal distribution ($n = 1$) to the exponential 
function ($n = \infty $). From actual fits in $I$ and $K'$ de Grijs {\it 
et al.} (1997) found (Fig. 1)
\begin{equation}
{2 \O n} = 0.54 \pm 0.20.
\end{equation}
I will take the whole range from the sech-function
to the exponential (that is $n=2 - \infty $; $ 2/n = 0-1$)
into account in what follows, which is reflected in
the ``uncertainties'' in the coefficients in the equations below.

\begin{figure}
\centerline{\vbox{
\psfig{figure=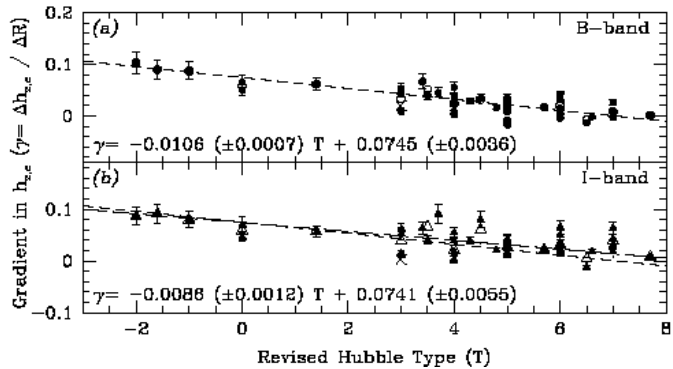,width=9cm}
}}
\caption{The radial change in the vertical scaleheight in a sample of 
edge-on galaxies (change in kpc in $h\subr{z}$ per kpc in $R$) as a 
function of morphological type. Except for the earliest types (S0 to
Sab) the change is very small (from de Grijs \&\ Peletier, 1997).}
\end{figure}

The constancy of the vertical scaleheight $h\subr{z}$ with 
radius has been studied in detail by de Grijs \&\ Peletier (1997) (see Fig. 2)
in the optical and near-IR and been confirmed at least for late-type galaxies.
Disks in early type disks might have a small variation of $h\subr{z}$ 
with radius.

With a constant mass-to-light ratio $M/L$, the luminosity density $L(R,z)$ 
is proportional to the space density $\rho (R,z)$.
Then the surface density becomes
\begin{equation}
\Sigma (R) = (2.6 \pm 0.6)\rho (R,0) h\subr{z},
\end{equation}
the vertical velocity dispersion in the plane
\begin{equation}
\left( \sigma \subr{z} \right) \subr{z=\circ } = \sqrt{(4.0 \pm 0.9) G 
\Sigma (R) h\subr{z}},
\end{equation}
and the z-velocity dispersion integrated perpendicular
to the plane
\begin{equation}
\sigma \subr{z} = \sqrt{(5.0 \pm 0.2) G \Sigma (R) h\subr{z}}.
\end{equation}

For the vertical velocity dispersion of the stars we expect for a constant
$h\subr{z}$
\begin{equation}
\sigma ^{2}\subr{z} \propto {\rm \;e}^{-R/2h\subr{R}}.
\end{equation}
This is consistent with observations by van der Kruit \&\ Freeman
(1986) and in Roelof Bottema's thesis (1995, see also 1993), at least out 
to about 2 scalelengths (in $B$). Furthermore,
Rob Swaters in his thesis (1999; chapter 7) found it also in the late-type
dwarf UGC 4325, again out to about 2 scalelengths.
On the other hand, Gerssen {\it et al.} (1997) could not confirm
it in NGC 488: they suggest that the scalelength in $B$ is not 
representative for the mass distribution. Also, NGC 488 is of early 
type and may not have a constant $h\subr{z}$.
So, the verification of eq. (7) is not complete. 
It is important to obtain $K$-band surface photometry
for NGC 488 and the other galaxies in Bottema's sample.

\begin{figure}
\centerline{\vbox{
\psfig{figure=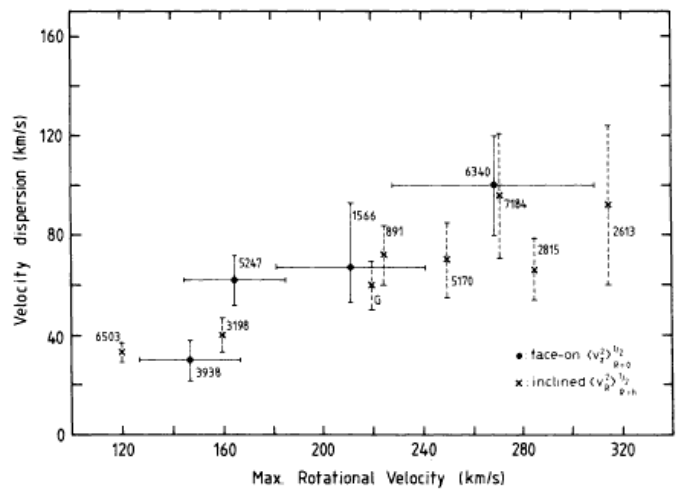,width=9cm}
}}
\caption{The relation between the stellar velocity dispersion and the 
maximum rotation velocity in disks of galaxies. The velocity dispersion
is the vertical one at the center for face-on
systems and the radial one at one scalelength in highly inclined systems
(from Bottema, 1993).}
\end{figure}

Bottema also found from a sample of 12 galaxies that the (extrapolated)
vertical dispersion at the center and the radial one at one scalelength
\begin{equation}
\sigma \subr{z,0} \sim \sigma \subr{R,h\subr{R}} = (0.29 \pm 0.10) V\subr{max},
\end{equation}
where $V\subr{max}$ is the rotation velocity in the
flat part of the rotation curve (Fig. 3).
Even the late-type dwarf UGC 4325 was found to follow this relation
(Swaters, 1999; $\sigma \subr{R,h\subr{R}} 
\sim 20$ km/s, $V\subr{max} \sim 90$ km/s).
It probably arises as follows:

Take the equation for the Toomre (1964) parameter $Q$ for local
stability
\begin{equation}
Q = {{\sigma \subr{R} \kappa } \O {3.36 G \Sigma }}.
\end{equation}
Use a Tully-Fisher relation $L \propto V\subr{max}^{n}$
with $n = 4$
and note that for a flat rotation curve the epicyclic frequency
$\kappa = \sqrt{2}V\subr{max}/R$. Then (9) can be rewritten as
\begin{equation}
\sigma \subr{R,h\subr{R}} \propto Q \left( {M \O L} \right) \sqrt{\mu 
\subr{\circ}}\ V\subr{max}.
\end{equation}
Here $\mu \subr{\circ}$ is the central face-on surface
brightness in $L\subr{\odot} {\rm pc}^{-2}$; $\mu \subr{\circ}$ and 
$M/L$ refer to the old disk population.
So the Bottema relation implies that 
\begin{equation}
Q \left( {M \O L} \right) \sqrt{\mu \subr{\circ}} \sim {\rm \; constant},
\end{equation}
even for low surface brightness dwarfs.
Observations of course allow a substantial scatter in this quotient
(or the individual terms) among galaxies.

For ``normal'' disks obeying ``Freeman's law'' $\mu \subr{\circ}$ 
translates into 21.7 B-mag arcsec$^{-2}$ and this then implies 
\begin{equation}
Q \left( {M \O L} \right) \subr{B} \sim 6.
\end{equation}

If we ignore for the moment the (dynamical) influence of the gas, it
can be shown (van der Kruit \&\ de Grijs, 1999) that at 
$R = 1h\subr{R}$
\begin{equation}
\sigma \subr{R,h\subr{R}} = (0.48 \pm 0.02) Q {{\sigma 
\subr{z,h\subr{R}}^{2}} \O {V\subr{max}}} {h\subr{R} \O h\subr{z}}.
\end{equation}
Here again the ``uncertainty'' in the coefficient relates to the
range of vertical density distributions above.
With the Bottema relation (8) this reduces to
\begin{equation}
\left( {{\sigma \subr{z}} \O {\sigma \subr{R}}} \right) ^{2}\subr{R=h\subr{R}} =
{{(7.2 \pm 2.5)} \O Q} {h\subr{z} \O h\subr{R}}.
\end{equation}
In the solar neighborhood the axis ratio of the velocity ellipsoid
$\sigma \subr{z} / \sigma \subr{R} \sim 0.5$
(Dehnen \&\ Binney, 1998).
If this also holds at $R=1h\subr{R}$ and using
$h\subr{z} \sim 0.35$ kpc and $h\subr{R} \sim 4$ kpc, it follows that
\begin{equation}
Q \sim 2.5.
\end{equation}

The HI-layer thickness can be used to estimate the disk surface density.
HI  observations of face-on galaxies (e.g. van der Kruit \&\ 
Shostak, 1984) indicate an HI velocity dispersion $\sigma \subr{HI}
= 8 - 10$ km s$^{-1}$. We can write (van der Kruit, 1981)
\begin{equation}
\left( {\rm FWHM} \right) \subr{HI} = (2.8 \pm 0.2) \sigma \subr{HI} 
\sqrt{{h\subr{z} \O {2 \pi G \Sigma (R)}}}.
\end{equation}

Another estimate of the disk mass $M\subr{D}$
follows from the global stability
criterion of Efstathiou {\it et al.} (1982)
\begin{equation}
Y = V\subr{max} \sqrt{{h\subr{R} \O {G M\subr{D}}}} \sim 1.1.
\end{equation}

For a pure exponential disk the maximum in the rotation
curve (Freeman, 1970) occurs at $R \sim 2.2 h\subr{R}$ with 
an amplitude
\begin{equation}
V\subr{max}^{\rm disk} = 0.88 \sqrt{\pi G \Sigma (0) h\subr{R}}.
\end{equation}
Hydrostatic equilibrium at the center gives
\begin{equation}
\sigma \subr{z,0}^{2} = (5.0 \pm 0.2) G \Sigma (0) h\subr{z}.
\end{equation}
Eliminating $\Sigma (0)$ between (18) and (19) and using the Bottema
relation (8) gives
\begin{equation}
{{V\subr{max}^{\rm disk}} \O V\subr{max}} = (0.21 \pm 0.08) \sqrt{h\subr{R} \O 
h\subr{z}}.
\end{equation}
So, the flattening of the disk can be used to test the ``maximum disk
hypothesis'', for which fits to rotation curves usually give
about 0.8 to 0.9 for this ratio.

\begin{figure}[t]
\centerline{\vbox{
\psfig{figure=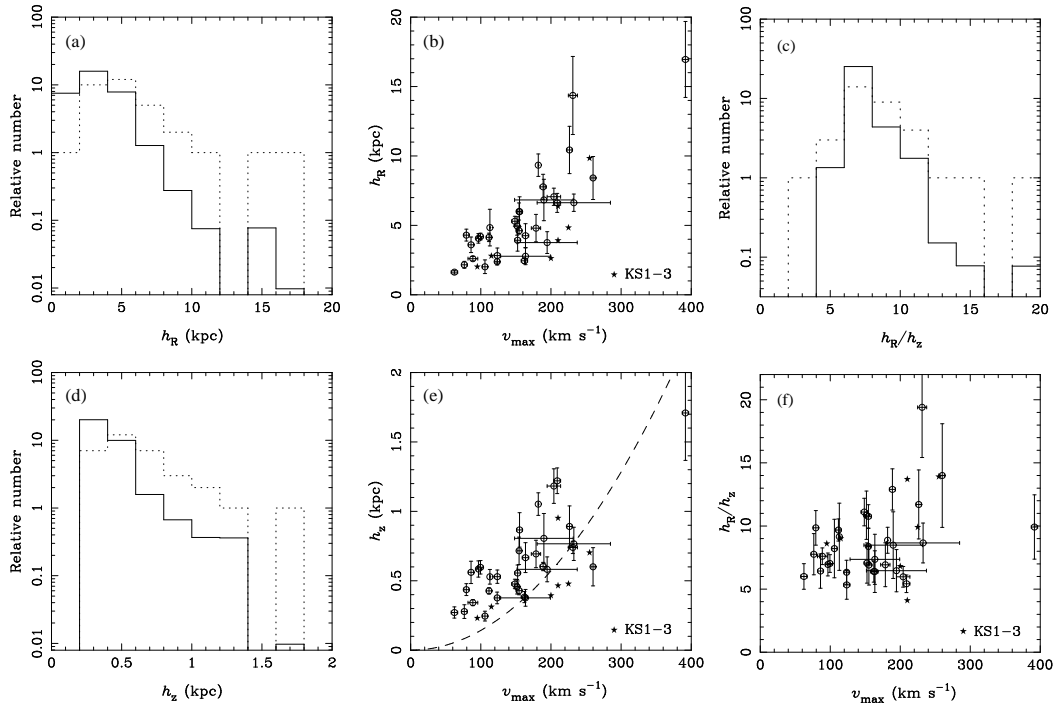,width=14cm}
}}
\caption{The radial scalelength $h\subr{R}$ and vertical
scaleheight $h\subr{z}$ for a sample of 34 edge-on galaxies. Panels (a) and
(b) show the distribution (solid line is corrected for sample selection) 
of $h\subr{R}$ and the correlation with the rotation velocity. Panels (d) 
and (e) show the same for $h\subr{z}$ and (c) and (f) for the ratio
$h\subr{R}/h\subr{z}$ (after Kregel {\it et al.}, 2001).}
\end{figure}

\section{Observations of disks}

The most extensive study of the photometric disk parameters in
the optical and near-IR is that of
a statistically complete sample of 86 disk dominated galaxies in
Roelof de Jong's thesis (1995, 1996abc; de Jong \&\ van der Kruit, 1994).
Some of his conclusions are:\\
$\bullet $ Freeman's law is really an upper limit to the central surface 
brightness.\\
$\bullet $ The scalelength $h\subr{R}$ does not correlate with Hubble type.\\
$\bullet $ In disks fainter regions are generally bluer, probably
resulting from a combination of stellar age and metallicity gradients.\\
$\bullet $ Outer regions are on average younger and of lower abundance.

\begin{figure}
\centerline{\vbox{
\psfig{figure=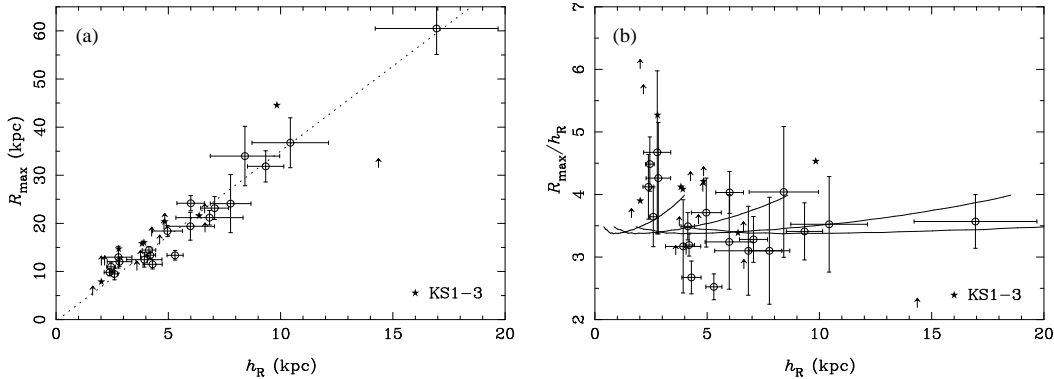,width=14cm}
}}
\caption{The relation between the truncation radius $R\subr{max}$ and
the radial scalelength $h\subr{R}$ of the disks in a sample of edge-on 
galaxies. The
righthand panel shows the ratio $R\subr{max}/h\subr{R}$ as a function of 
$h\subr{R}$ and
the lines are models based on the formalism of Dalcanton {\it et al.}, 1997 
(from Kregel {\it et al.}, 2001).}
\end{figure}

Richard de Grijs (1997, 1998; de Grijs \&\ van der Kruit, 1996) presented 
optical and near-IR surface photometry of a sample of 47 edge-on galaxies.
The data have recently been re-analysed by Kregel {\it et al.} (2001;
see also de Grijs {\it et al.}, 2001) with a new and improved 
2-D fitting procedure. Some results (Fig. 4 and 5):\\
$\bullet $ Both $h\subr{R}$ and $h\subr{z}$ correlate in general terms
with $V\subr{max}$.\\
$\bullet $ For $h\subr{R}$ this is expected from the Tully-Fisher 
relation.\\
$\bullet $ Our Galaxy would be somewhat unusual if the scalelength $h\subr{R}$
is as small as 2 to 2.5 kpc as some recent studies claim (see also van
der Kruit, 2000).\\
$\bullet $ The flattening of the sample after volume correction is
${h\subr{R}/h\subr{z}} = 7.3 \pm 2.2$. Then 
\begin{equation}
{V^{\rm disk}\subr{max} \O V\subr{max}} = 0.57 \pm 0.22.
\end{equation}
So most galaxies appear not to be ``maximum disk''. Recall that this 
result follows directly from the observations using only the rotation 
curve of the self-gravitating exponential disk, hydrostatic equilibrium
and Bottema's (empirical, but explainable) relation  (8).
Bottema (1993) derived a similar result in the analysis of his sample
of galaxies, in which he measured the stellar kinematics and found 
$V^{\rm disk}\subr{max}/V\subr{max} = 0.63 \pm 0.17$.\\
$\bullet $ At least 20 of the spirals show radial truncations.\\
$\bullet $ The ratio of the truncation radius and the scalelength is
${R\subr{max}/h\subr{R}} = 3.6 \pm 0.6$.\\
$\bullet $ But  large galaxies have smaller values for this ratio than 
small ones.\\
$\bullet $ For common disks with scalelengths of 5 kpc or less, the ratio
is about 4.

The truncation radius in a simple view results from the maximum
specific angular momentum of the sphere from which the disk collapsed.
Van der Kruit (1987), in the context the Fall \&\ Efstathiou (1980) 
picture of disk galaxy formation, then predicted a value of 4.5
for the ratio, based on a Peebles (1971) spin parameter 
$\lambda = J |E|^{1/2} G^{-1} M^{-5/2}$ of 0.7.
Dalcanton {\it et al.} (1997) have extended this to a models with a dispersion
in the spin parameter.
We have calculated model surface density profiles with their 
method for $M\subr{tot} = 10^{10} - 10^{13} M\subr{\odot}$ and 
$\lambda = 0.01 - 0.28$. These are the lines in Fig. 5b.

For completeness I mention that in many cases there is a warp in the
HI-layer in the outer parts, often starting roughly at the truncation radius.
This suggests that the warp material has been accreted subsequent
to disk formation.

\section{IC 5249}

The ``superthin'' disk galaxy IC 5249 has recently been studied by van der Kruit
{\it et al.} (2001). In this paper, we re-analyse the ATCA HI 
observations reported in  Abe {\it et al.} (1999), 
use the photometry of three Ken Freeman students, Claude Carignan (1983), 
Richard Wainscoat (1986) and Yong-Ik Byun (1992, 1998),
and present a measurement of the stellar kinematics.

The HI observations show --contrary to the conclusions of Abe {\it et
al.}-- a rotation curve that is flat over a large part of the disk at 
$V\subr{max} = 105 \pm 5$ km s$^{-1}$. From the available photometry we
adopt $h\subr{R} = 7 \pm 1$ kpc, $h\subr{z} = 0.65 \pm 0.05$ kpc, 
$\mu \subr{\circ} \sim 24.5$ B-mag arcsec$^{-2}$ and
$R\subr{max} = 17 \pm 1$ kpc.
Note that $h\subr{z}$ is larger than that in our Galaxy!
It follows that 
IC 5249 appears on the sky as a very thin disk, because it combines
a low surface brightness with a very long scalelength and a truncation
radius after only about 2.5 scalelengths.

From our data we can derive various kinematical and dynamical
properties, using the equations given above.
\begin{table}
\caption{Properties of IC 5249 (top part observed, lower part derived)}
\begin{tabular}{||l|c|c||}
\hline
 & $R = 7$ kpc & $R = 17$ kpc \\
\hline
$\sigma \subr{\theta }$ (km/s) & 25-30 & -- \\
$V\subr{rot}$ (km/s) & $90 \pm 5$ & $105 \pm 5$ \\
$dV\subr{rot}/dR$ (km/s.kpc) & $3 \pm 4$ & $ 0 \pm 1$ \\
$\kappa $ (km/s.kpc) & $20 \pm 6$ & $3.5 \pm 0.4$ \\
$\sigma \subr{R}$ (km/s) & $35 \pm 5$ &  -- \\
Asym. drift (km/s) & $10 \pm 3$ & -- \\
\hline
$\sigma \subr{R}$ (km/s) & -- & $25 \pm 5$ \\
$\Sigma (M\subr{\odot}$ pc$^{-2}$) & $ \sim 25$ & $\sim 6$ \\
$\sigma \subr{z}$ (km/s) & $ 19 \pm 4 $ & $11 \pm 2$ \\
$Q$ & $\sim 2$ & $\sim 2$ \\
(FWHM)$\subr{HI}$ (kpc) & $0.60 \pm 0.17 $ & $1.5 \pm 0.5$ \\
\hline
\end{tabular}
\end{table}

At $R = 7$ kpc the stellar velocity dispersions
are similar to the solar neighborhood, while 
the surface density of the disk is about half that of the solar neighborhood.
Star formation must have proceeded much slower in IC 5249 in order
to give the low surface brightness.
But, surprisingly, as much dynamical evolution has occured as in the 
Galactic disk. Also note that with $\sigma \subr{R,h}/V\subr{max} = 0.33 
\pm 0.05$, IC 5249 falls on the Bottema relation (8).

\acknowledgments
I am grateful to the Mount Stromlo Observatory and John Norris for hospitality
during the preparation of this review. Michiel Kregel kindly allowed me to 
quote from our recent paper and provided Fig.s 4 and 5. I thank Ken Freeman 
for stimulating discussions and collaborations over many years and wish 
him well.

\section*{Discussion}

\noindent {\it King:\, } One of the tightest correlations you showed was
between $R\subr{max}$ and $h\subr{R}$. What does $R\subr{max}$ mean, and
is there any obvious reason for such a tight correlation?

\noindent {\it van der Kruit:\, } $R\subr{max}$ is the truncation radius.
The tightness of the correlation results from the fact that over
scalelengths in the range from 1 to 15 kpc or so, the truncation radius
occurs at 3 -- 4 scalelengths. I have commented on this in the paper
above.

\noindent {\it Byun:\, } In the 1970 Freeman paper, disk galaxies were
divided into two groups, type I and type II. Type II galaxies had
exponential disks which {\it do not} continue into the centers. What are
our current understandings of these type II galaxies?

\noindent {\it van der Kruit:\, } Although there still are some systems
having profiles that were called type II, it appears that these no
longer make up a substantial fraction of the observed profiles. The
prime example, M83, is barred and the bar may be the cause of this
behavior, but there are also galaxies that are unbarred and display the
type II characteristic. But again it no longer is an important class in
modern surface photometry profiles.

\noindent {\it Fall:\, } My impression is that realistic angular momentum
distributions $M(h)$ produced by hierarchical clustering do not have the sharp
features that would lead to edges in the disks. My preferred explanation
for the edges at $R\subr{edge} \approx  4 r\subr{disk}$ is in terms of
local instabilities in the disks (Goldreich \& Lynden-Bell and
Toomre-type instabilities\footnote{The references to the papers by 
Goldreich \&\ Lynden-Bell (1965) and Toomre (1964) appear in Table 1 
[PCvdK]}). George Efstathiou and I showed these would
give rise to $R\subr{edge} \approx  4 r\subr{disk}$.

\noindent {\it van der Kruit:\, } I am aware of the Fall \&\ Efstathiou (1980)
explanation. Actually, in my third paper with Leonard Searle on edge-on galaxies
(\aap, 110, 61, 1981) we refer to this and find that it predicts truncations
at $0.8 \pm 0.2$ times the observed ones for our sample.
 
\noindent {\it van der Kruit:\, } After the session Mike Fall suggested
that the conclusion that the maximum disk hypothesis does not apply to
most galaxies using the disk flattening may be affected by underestimating
the scaleheight due to the presence of stellar generations in the disks 
with a range in vertical velocity dispersions and scaleheights. The younger of 
these are brighter and their smaller dispersions lead to a lower scaleheight
in the photometry than in the mass distribution. It is well-known that the 
observed diffusion of stellar random motions gives --according to 
observations in the solar neighborhood-- a velocity dispersion 
proportional to $\sqrt{\rm age}$  and Mike points out that the
stellar generations therefore have a scaleheight proportional 
to age.  I was urged by some participants to reply to this. \\
According to eq. (20) a systematic underestimate of the scaleheight
$h\subr{z}$ results in an {\it overestimate} in the ratio $V\subr{max}^{\rm
disk}/V\subr{max}$ and allowing for it then takes the disks even 
further from maximum disk. However, the effect must also
occur for the observed velocity disperions in face-on disks, 
so Bottema would systematically have
underestimated the dispersions resulting in an underestimate of the
coefficient in eq. (8) and (20). I have returned to some old notes to myself on
this and updated these. In view of the potential importance of the
effect I document this here.\\
The luminosity of a generation of
stars as a function of age can be estimated in three ways. First look at
the luminosity of the main sequence (MS) turn-off stars. The stellar MS 
luminosity is roughly proportional to $M^{3}$ and MS lifetime $\tau $ to
$M^{-2}$. So luminosities $L\subr{MS}$ are roughly proportional to 
$\tau ^{-3/2}$. Secondly, we
may look at the giants only. I estimate from evolutionary tracks that at
the tip of the giant branch the luminosity $L\subr{Giant}$ is crudely
proportional to the mass on the MS (at least for stars between one and a few 
solar masses; MS lifetimes of 1 to 10 Gyr), so that $L\subr{Giant} \propto \tau
^{-1/2}$. Finally, single burst models of photometric evolution can be
used. A nice example is illustrated in Binney \&\ Merrifield (Galactic
Structure; Princeton Univ. Press, 1998) in Fig. 5.19 on page 318. The
$(M/L)_{B}$ is proportional to age, so that the luminosity of a single
burst $L\subr{SB} \propto \tau ^{-1}$. This is --as expected-- nicely in
between $L\subr{MS}$ and $L\subr{Giant}$.\\
To estimate the effect it is most practical to look at the integrated
velocity dispersion in a face-on disk. The weighted velocity dispersion 
can then be estimated by integrating over the relevant ages $\tau $ in
\begin{displaymath}
\langle \sigma \subr{z}^{2} \rangle = {{\int {\rm SFR}(\tau ) L(\tau ) 
\sigma ^{2}(\tau ) d\tau } \O {\int {\rm SFR}(\tau ) L(\tau ) d\tau }}.
\end{displaymath}
To estimate the error this should then be compared to the case of equal
weighing of the generations
\begin{displaymath}
\langle \sigma \subr{z}^{2} \rangle = {{\int {\rm SFR}(\tau ) 
\sigma ^{2}(\tau ) d\tau } \O {\int {\rm SFR}(\tau ) d\tau }}.
\end{displaymath}
We had $\sigma ^{2} \propto \tau $ and may take for late type disks as a
reasonable approximation a constant star formation rate SFR$(\tau )$.\\
In order to check the effect on the scaleheights the best approximation is
to use $L\subr{SB}$. The results depend on the range in $\tau $ we take to 
perform the integration. Since fits for $h\subr{z}$ are made above the 
dust lanes we need to ignore the youngest generations. As examples I
take the integration from 2 and 3 to 10 Gyr ; this means ignoring generations
in which the scaleheights are
less than 0.2 and 0.3 times that of the oldest generations. The values for
$\sqrt{\langle \sigma \rangle ^{2}}$ are then underestimated by
respectively 9 and 6\%\ and the scaleheights by 17 and 11\%.\\
For the effect on the velocity dispersions we have to take
$L\subr{Giant}$, since thes are measured by comparing to a late-GIII or 
early-KIII template star. The galaxy spectra refer in practice for this 
mostly to interarm regions, so it
is fair to take the lower limit in the integration now at 1 or 2 Gyr.
Then $\sqrt{\langle \sigma \rangle ^{2}}$ is underestimated by
respectively 8 and 4\%. Even if the lower integration limit is set to zero
(assuming red supergiants fit the luminosity class III template star), 
the error is only 18\%.\\
So, the effects are small and furthermore in an application of eq. (20)
we need to take the square root of the 
scaleheight, while the two effects work in opposite directions.

\end{document}